\documentclass[10pt]{iopart}
\usepackage{dsfont}
\usepackage[T1]{fontenc}
\usepackage[T1]{url} 
\expandafter\let\csname equation*\endcsname\relax
\expandafter\let\csname endequation*\endcsname\relax
\usepackage{amsmath, amssymb, graphicx, booktabs, textcomp, mathabx, cite, braket}
\usepackage{times}
\usepackage[hyperindex,
            colorlinks=true,linktocpage=false,
            bookmarks=true,
            pdftitle={Bayesian inference on EMRI signals using low frequency approximations},
            pdfauthor={Asad Ali},
            pdfkeywords={Bayesian Analysis, EMRI, MCMC, Parallel Tempering MCMC}]{hyperref}
\hypersetup{
	bookmarksnumbered,
	pdfstartview={FitH},
  	citecolor={black},
  	linkcolor={black},
  	urlcolor={black},
	pdfpagemode={UseOutlines}
}
\usepackage[nodisplayskipstretch]{setspace}
\begin{document}
\urlstyle{sf}
 \title{Bayesian inference on EMRI signals using low frequency approximations}

\author{Asad Ali$^{1,2}$,
         Nelson Christensen$^3$,
         Renate Meyer$^1$, and 
         Christian R\"{o}ver$^{4}$\footnote{Present address: Department of Medical Statistics, University Medical Center G\"{o}ttingen, 37073~G\"{o}ttingen, Germany.}
}
\address{$^1$ Department of Statistics, The University of Auckland,
         Auckland, New Zealand}
\address{$^2$ Department of Space Science, Institute of Space Technology (IST), Islamabad
44000, Pakistan}
\address{$^3$ Department of Physics \& Astronomy, Carleton College,
         Northfield, MN, USA}
\address{$^4$ Max-Planck-Institut f\"{u}r Gravitationsphysik (Albert-Einstein-Institut) and Leibniz Universit\"{a}t Hannover, 30167 Hannover, Germany}

\eads{\mailto{asad.ali@ist.edu.pk}, 
      \mailto{nchriste@carleton.edu}, 
      \mailto{meyer@stat.auckland.ac.nz}, and 
      \mailto{christian.roever@med.uni-goettingen.de}}
 
\begin{abstract}\small
Extreme mass ratio inspirals (EMRIs) are thought to be one of the most exciting
gravitational wave sources to be detected with LISA. Due to their complicated
nature and weak amplitudes the detection and parameter estimation of such
sources is a challenging task. In this paper we present a statistical methodology
based on Bayesian inference in which the estimation of parameters is carried out
by advanced Markov chain Monte Carlo (MCMC) algorithms such as parallel
tempering MCMC. We analysed high and medium mass EMRI systems that
fall well inside the low frequency range of LISA. In the context of the Mock
LISA Data Challenges, our investigation and results are also the first instance
in which a fully Markovian algorithm is applied for EMRI searches. Results
show that our algorithm worked well in recovering EMRI signals from different
(simulated) LISA data sets having single and multiple EMRI sources and holds
great promise for posterior computation under more realistic conditions. The
search and estimation methods presented in this paper are general in their
nature, and can be applied in any other scenario such as AdLIGO, AdVIRGO
and Einstein Telescope with their respective response functions.
\end{abstract}

\pacs{04.80.Nn, 02.70.Uu.}

\submitto{Classical and Quantum Gravity}

\section{Introduction}
It is likely that most of the galaxies, including our own milky way, host super massive black holes (SMBHs) with masses of order $10^6 M_\odot$ ($M_\odot=$ solar mass) or larger in their centres. These SMBHs are surrounded by a large population of stellar mass compact objects (COs) such as neutron stars, white dwarfs and small black holes with masses $\sim10 M_\odot$. Due to multi-body interactions most of these COs occasionally end up being trapped in an orbit which passes too close to the central mass. Once captured in the strong gravitational field of an SMBH, a CO then starts orbiting around the central mass in an eccentric orbit which decays with time and the CO eventually spirals into the central mass. These inspirals are called extreme mass ratio inspirals (EMRIs) because of the large difference in the masses of the two bodies.
EMRIs are considered to be one of the most important potential sources of gravitational waves
(GW's), to be detected with the laser interferometer space antenna\footnote{We realize that the state of LISA is in limbo after NASA's decision to leave the project, but the European Space Agency is investigating carrying on with a scaled down version of the project, and the conclusions we reach in this paper will certainly be applicable to that mission if it moves forward.} (LISA) \cite{Danzmann2003}. These inspirals may encode information about the structure of the central black hole, its evolution and other features such as the Lense\---Thirring effects \cite{Everitt2011} and spin\---orbit coupling \cite{Alberto2004}.

Bayesian approaches along with MCMC methods have been in use by different groups
working on the GW source detection and parameter estimation problems (see e.g. \cite{Nelson1998, Nelson01, Christian2007, ChristPhD} and
many others). In the context of EMRI signals the reversible jump MCMC algorithm was
employed in \cite{stroeer2006}. Another Monte Carlo, but not rigorously Markovian, approach in which the successive states are chosen from directed proposal distributions, was used in \cite{babak2009, gair2008, cornish2008}.

In this work we present results of the applications of the Bayesian approach using
the parallel tempering MCMC (PTMCMC) algorithm \cite{geyer} for the detection and parameter
estimation of EMRI signals in LISA data. The noise spectrum is assumed to be unknown
and for this reason we used Whittle's approximation to the Gaussian likelihood \cite{whittle1952}. The
Whittle likelihood uses the approximate properties of discrete Fourier transform (DFT) and
assumes that the DFTs are approximately independent normally distributed with mean zero
and variance proportional to the power spectral density.

This paper is organized as follows: in section 2 we present the Bayesian statistical model,
comprising the waveform model, our assumptions for the error distribution with unknown
power spectrum, and prior distributions of the parameters. Section 3 briefly explains the
Bayesian approach to detection and parameter estimation and details the MCMC search
algorithm. The implementation of the algorithm is described in detail in section 4. Conclusions
are given in section 5.

\section{Bayesian statistical model}
In this section, we outline all components of our statistical model that encompasses the
waveform model and detector response, the observation error assumptions that lead to the
Whittle likelihood and the prior distribution of the parameters.

\subsection{The waveform model and detector response}
The EMRI sources given in MLDC \cite{overview2} LISA data are generated by Barack and Cutler's
\textit{analytic kludge} waveform (AKW) approximation \cite{Barack}. The waveform model is described by
14-dimensional parameter set $\theta = (\nu_0, M, \mu, e_0, \tilde{\gamma}_0, \Phi_0, \alpha_0, \theta_S, \phi_S, \lambda, \chi, \theta_K, \phi_K, D_L)$, where $\nu_0$ is the initial orbital frequency, $\mu$ and $M$ are the masses of CO and SMBH respectively, $e_0$ is the initial eccentricity, $\tilde{\gamma}_0$, $\Phi_0$ and $\alpha_0$ are the initial orbital phase angles, $\theta_S$ and $\phi_S$ are the ecliptic latitude and longitude, respectively, $\lambda$ is the orbital inclination, $\chi$ is SMBH's spin with $\theta_K$, $\phi_K$ its orientation angles and $D_L$ is luminosity distance. In the Barack and Cutler parametrization, the log-transformed values of the parameters $\nu_0$, $\mu$, $M$ and $D_L$ are used. For the MCMC searches a suitable alternative is to use the truncated AKWs \cite{gair2008}. These waveforms can be computed $\sim3$ times faster than full AKWs and there is a typical overlap of $\sim90\--95\%$ between the two. The multiple harmonics and the complex structure of EMRIs pose a challenge as far as their computation and the statistical estimation of their source
parameters is concerned. The likelihood surface for EMRIs contains multiple local peaks
corresponding to different harmonics, of which some are as high as 85\% of the peak of the
dominant harmonic. Furthermore, the orbital evolution (over time) introduces much more
uncertainty to the likelihood surface. This means that the orbital parameters keep changing
throughout the signal life. To overcome these issues, sophisticated algorithms are needed that
can deal with multi-modality due to multiple harmonics and the increased uncertainty due to
the time-varying nature of orbital parameters. One such algorithm will be presented in the
subsequent sections.

The full descriptions of both the full AKW and truncated AKW models are extensive so
we give only the final expressions of the model which is a pair of two polarization signals defined as
\begin{align}\label{equation1}
 h_+(t)      &= A^+ (t) \cos 2\psi(t) + A^\times(t) \sin 2\psi(t),\\
\label{equation2}
 h_\times(t) &=-A^+ (t) \sin 2\psi(t) + A^\times(t) \cos 2\psi(t),
\intertext{where $\psi(t)$ is the polarization angle defined as}
\label{equation3}
\psi(t) &= \arctan\left(
\dfrac{\cos \theta_S \sin\theta_L(t) \cos(\phi_S - \phi_L(t)) - \cos\theta_L(t) \sin\theta_S}{\sin \theta_L(t) \sin(\phi_S - \phi_L(t))}\right),
\end{align}
where $\theta_S$ and $\phi_S$ are ecliptic latitude and longitude, respectively, and $\theta_L(t)$ and $\phi_L(t)$ are the
time-varying angles specifying the instantaneous direction of the angular momentum.

Unlike ground-based interferometers the LISA arms are at $60^\degree$ angles, and the
interferometers are susceptible to '+' and '$\times$' polarizations. Due to its orbital dynamics,
the detector response is complicated. The detector output may be derived numerically
\cite{syntheticLISA, CornRubbo, LISASIM}, while in the limit of long wavelength/low frequency approximation (LFA) the
mapping simplifies \cite{Barack, Cutler1998}, which is the regime we will be concerned with here. The three 
spacecraft's output may be re-combined into two variables with stochastically independent
noise components, namely the '$A$' and '$E$' variables \cite{Prince}; data analysis in the following is
going to be based on these two time series.

Although the LFA is simpler and faster, the amplitude of the final LISA response is not
exactly the same as that of \cite{CornRubbo, LISASIM}, which we will refer to as the full LISA response throughout
this paper. While working on MLDC 4 blind data, we found that the amplitudes of the LFA
response are by a factor of $\sim3$ larger than those derived by full LISA response. We compared
the two responses for several (noise-less) EMRI signals posted on the MLDC webpage \cite{MLDC}
including those given in the training data of MLDC 4 and saw the same phenomena. This
difference in amplitudes results in an incorrect estimation of luminosity distance ($D_L$) and sky
location angles ($\theta_S$, $\phi_S$) as we have found in several test MCMC searches on MLDC 1B EMRI
data sets. Particularly, during MCMC searches the chains for sky location angles used to get
locked at wrong positions and then nothing could move them from those positions. The same
phenomenon (wrong estimation of distance and sky location) was observed by another group
too \cite{gair2008, GairPorter2008}. As an ad hoc solution we divided the overall amplitude in our signal model by
3 and used this option throughout the MLDC 4 blind searches, more investigation regarding
the exact origin is required. When the overall amplitude in LFA was divided by 3, the sampler
was able to converge to the correct values of the luminosity distance and sky location angles.
It is of interest to note that in both cases (i.e. the adjusted and unadjusted amplitudes) the other
key parameters, e.g. $\nu$, $\mu$, $M$, $e$ and $\chi$ are almost unaffected and remain the same. Later on,
near the submission of MLDC 4 blind entries, we further found that for high mass EMRIs the
difference factor is generally rather larger than 3, it is generally $\sim6$.

\subsection{Prior information}
In order to implement the a priori information about the parameters to be inferred, we followed
the specifications given in the MLDC Task Force's documentation \cite{overview2, MLDC}, and for different
parameters we employed the following prior distributions: $\mu \in \text{Uniform}[9.5, 10.5]M_\odot$,
$e_0\in\text{Uniform}[0.15, 0.25]$, $\chi\in\text{Uniform}[0.5, 0.7]$. The range of the prior distribution of $e_0$ was changed to $[0.10, 0.40]$ because of its initial values given for different MLDC EMRI sources being beyond the range $[0.15, 0.25]$. The prior for SMBH mass are different for different
sources: $M\in\text{Uniform}[0.95, 1.05] 10^7M_\odot$ (high mass), $M\in\text{Uniform}[4.75, 5.25] 10^6M_\odot$
(medium mass) and $M\in\text{Uniform}[0.95, 1.05] 10^6M_\odot$ (low mass). For initial azimuthal orbital frequency $\nu_0$, the prior distribution was set to be uniform over the range $[5.0 \times 10^{-5}, 0.01]$.
For the polar angles we assumed isotropic priors. For some parameters such as $\nu_0$, $\mu$, $M$
and $D_L$, their logarithmic values were used; as in \cite{Barack}, for ease of specification of proposal
distributions, the prior densities for these parameters were transformed accordingly. For the
noise power spectrum we used the conjugate Inv$-\chi^2$ prior distribution, where the prior scale
parameters were estimated based on a disjoint stretch of the same data set \cite{Christian2011}. In the following
subsections we present some results obtained with the application of the above search and
estimation algorithm on different MLDC data sets containing single and multiple EMRI
sources.

\section{Parameter estimation}
In this section, we briefly describe the Bayesian approach to statistical inference and posterior computation via MCMC. Furthermore, we give an overview of the tempering methods and the parallel tempering strategy that we employed.
\subsection{Posterior inference} 
We are interested in inferring a priori unknown parameters $\theta$ from measured data $y$. To
this end we derive the parameters' posterior probability distribution, which expresses the
information on the actual parameter values after consideration of the observed data by assigning
probabilities to regions of the parameter space. Given the relationship between parameters and
data through the likelihood function $p(y|\theta)$, the posterior density function is given by Bayes' theorem as
\begin{equation}\label{equation4}
p(\theta|y) = \dfrac{p(y|\theta) p(\theta)}{p(y)}
\end{equation}\cite{Gelman03}. 
Here $p(\theta)$ is the prior probability density, expressing any information we have about the
parameters before accounting for the measured data. The posterior distribution (\ref{equation4}) is essentially
given by the product of prior and likelihood, while the evidence 
$p(y) = \int p(y|\theta ) p(\theta ) d\theta$
is commonly of minor concern for parameter estimation purposes, as it only constitutes a normalizing constant.
\subsection{Monte Carlo integration}
Bayes theorem supplies the posterior probability distribution in terms of its density function.
In order to extract information relevant for any particular purpose, one may be interested
in marginal posterior distributions, posterior expectations, quantiles, etc; what is commonly
required is the evaluation of integrals with respect to the posterior distribution. As these
integrals are rarely analytically tractable, stochastic integration methods are commonly used to approach posterior inference. Such Monte Carlo methods aim at approximating the desired
integrals via random sampling; instead of computing an expectation value analytically, random
draws from the distribution of interest are generated, and the expectation is then approximated
by an average. Analogous procedures are applied for other figures of interest, like quantiles or
marginal densities for example, and an arbitrary accuracy may be achieved by increasing the
sample size. A popular variety of such Monte Carlo procedures is the Metropolis(\---Hastings)
algorithm \cite{Hastings70}; what makes this particular algorithm attractive for Bayesian inference is the
fact that it only requires the unnormalized probability density function of the distribution of
interest as an input, which in this case is supplied by Bayes' theorem (\ref{equation4}). Instead of providing
stochastically independent samples from the distribution, the Metropolis(\---Hastings) algorithm
will generate a Markov chain of subsequently dependent draws whose stationary distribution
is the distribution of interest. At each step in the generation of the random sequence, only ratios
of probability density values need to be considered, so that an overall normalization constant
(like $p(y)$ in (\ref{equation4})) does not need to be known. To approximate a target distribution with density
function $p(\theta|y)$, the Metropolis(\---Hastings) algorithm employs an acceptance-rejection rule
to construct a Markov chain from an auxiliary density $q(\cdot|\cdot)$, which is referred to as proposal
density or transition probability function. Being at current state $\theta^{(t)}$ the acceptance probability for moving to a new state $\theta^\prime$  is defined as
\begin{align}\label{equation5}
\alpha &= \min\left\lbrace 1, \dfrac{p(\theta^\prime|y)q(\theta^{(t)})|\theta^\prime
)}{p(\theta^{(t)}| y)q(\theta^\prime|\theta^{(t)})} \right\rbrace.
\intertext{Taking $q(\cdot|\cdot)$ to be symmetric, i.e. $q(\theta|\theta^\prime) = q(\theta^\prime|\theta)$, leads to the basic Metropolis algorithm \cite{Metropolis53} with acceptance probability given by}
\label{equation6}
\alpha &= \min\left\lbrace 1, \dfrac{p(\theta^\prime|y)}{p(\theta^{(t)}| y)} \right\rbrace.
\end{align}

\subsection{Tempering methods}
In most multi-dimensional cases, the density surface of complicated target distributions turns
out to have multiple secondaries or local modes that are well separated by deep valleys of low
probability regions. The simple Metropolis(\---Hastings) algorithm tends to get stuck at some of
these local modes for a prohibitively long time before reaching the global mode. Analogous
to an annealing process, temperature is used to scale the target density in order to flatten the
local modes so that the MCMC sampler moves freely towards the global maximum without
being trapped in local modes. At a given temperature $T \ge 1$, samples are generated from a
tempered version of the target density $p(\theta|y)$ defined as
\begin{equation}\label{equation7}
p_T (\theta|y) \propto p(\theta|y)^{\frac{1}{T}},
\end{equation}
where $T = 1$ yields the actual target distribution. This heating is equivalent to increasing the standard deviation of the target density by a factor $\sqrt{T}$, therefore as $T$ increases the heated distribution becomes flatter and gets closer to the uniform distribution, which enables the 
Markov chain to move more freely and hence faster towards higher probability regions.

Some of the most popular tempering methods are simulated annealing \cite{Kirkpatrick1983}, simulated
tempering \cite{Marinari1992} and PTMCMC aka Metropolis coupled-MCMC. PTMCMC introduced in \cite{geyer}, is a powerful optimization of the simple Metropolis(\---Hastings) algorithm which is very
effective in improving the mixing of MCMC chains and in particular in escaping the local
modes. The algorithm works by running multiple MCMC chains in parallel, each simulating
a separate target density characterized by a different temperature and occasionally attempting
swaps of its current states. As stated above, in principle, a high temperature chain sees the density of the target distribution as more flattened relative to a low temperature chain,
which means that the high temperature chain can move more freely across the valleys of low
probability regions in between modes. In order to make the low temperature chain benefit
from the high temperature chain that may be sampling near another mode, an exchange of
information about the current states is attempted by proposing a swap of the current states using
an additional Metropolis acceptance\---rejection step. Detailed descriptions of this algorithm can
be found in \cite{ChristPhD, gilks1996, Liu01}.

\subsection{The likelihood function}
The data, which is sampled at discrete time steps, can be represented by the following signal
plus noise model,
\begin{equation}\label{equation8}
y(t) = s(t, \theta) + \epsilon(t) \qquad t = 1, . . . ,N,
\end{equation}
where the deterministic component $s(t, \theta)$ is the true signal model depending on parameters
$\theta$ and $\epsilon(t)$ is a zero-mean stationary time series with unknown spectral density \cite{Christian2011}.

When the data are in (discrete) frequency domain for an unknown noise spectral density
the likelihood function is defined as
\begin{align}
 \label{equation9}
p(\tilde{y}|\theta) &= K \times \exp \left[ - \sum_{j=0}^{\nu} \left( \log(S(f_j)) +\sum_{j=1}^{\nu}\frac{|\tilde{y}(f_{j}) -
\tilde{s}(f_{j},\theta)|^2}{S(f_j)} \right)\right],  
\end{align}
where $\nu = \lfloor (N-1)/2 \rfloor$ is the greatest integer less than or equal to $(N-1)/2$, $\tilde{y}(f_j)$ and $\tilde{s}(f_j, \theta)$ are Fourier transformed observables and model signal, respectively, $S(f_j)$ is the one sided
power spectral density and $K$ is the normalizing constant.

When the noise spectrum is assumed known then one can omit the constant term, $\log(S(f_j))$, from equation (\ref{equation9}) to obtain
\begin{align} \label{equation10}
p(\tilde{y}|\theta) &= K \times \exp\left[-\sum_{j=1}^{\nu}
{\frac{|\tilde{y}(f_{j}) -
\tilde{s}(f_{j},\theta)|^2}{S(f_{j})}}\right].
\end{align}
In the literature, equation (\ref{equation9}) is known as Whittle's approximation to the Gaussian likelihood
or simply Whittle likelihood \cite{whittle1957}. The Whittle likelihood assumes that the DFTed residuals are approximately independent complex normally distributed with mean zero and power spectrum $S(f_j)$. The complete description of the Bayesian estimation of the noise spectral density is explained in \cite{Christian2011}. The likelihood computation is commonly simplified by restricting the summation to the limited frequency range relevant to the signals of interest. Furthermore, as mentioned in section 2, the two TDI observables $A$ and $E$ are independent, their joint likelihood is just the product of their individual likelihoods.

\section{Implementation}
Although LFA is much more time efficient compared to full LISA response, EMRI waveforms
are still computationally expensive and hence we had to use only shorter data segments
ranging in length from one to four weeks. We used heavy-tailed Student-t proposals which
are very useful for good mixing. The parallel computation on multiple processors or cores
is accomplished by including the message passing interface library \cite{Gropp1998MPI} with the relevant additional programming scripts in the main analysis code. In all of our searches we used eight or ten chain PTMCMC.\\

\subsection{Single EMRI sources: example searches}
\begin{figure}[!t]
 \centering
 \includegraphics[width=1.0\textwidth]{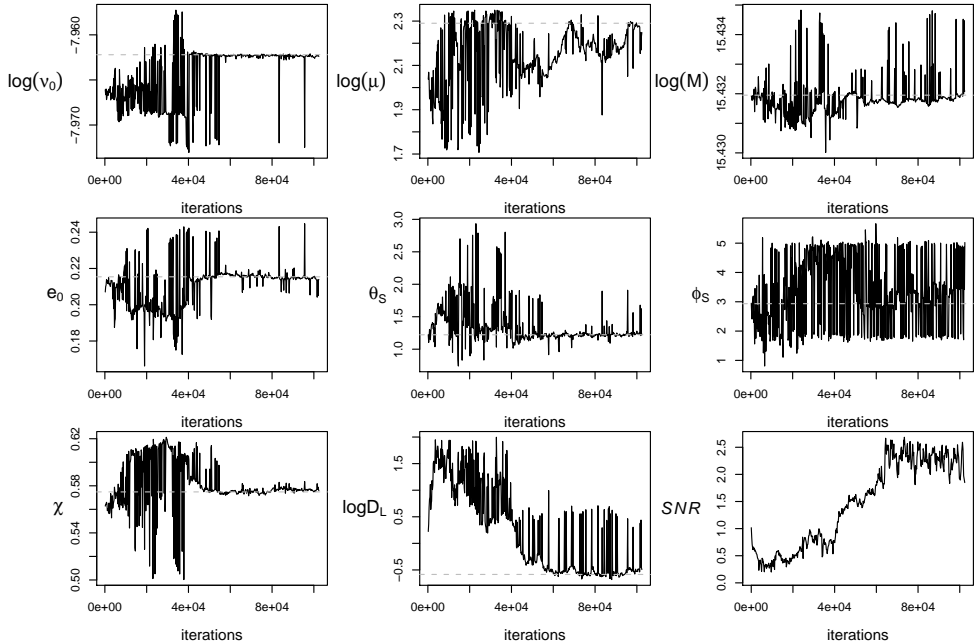} \centering
\caption{\label{Figure1} A typical (initial) shorter MCMC run. The trace plots of marginal posterior MCMC samples for some key parameters for the EMRI training source MLDC 1B.3.2. The grey dashed lines indicate the true values.}
\end{figure}

\begin{figure}[!h]
 \centering
 \includegraphics[width=1.00\textwidth]{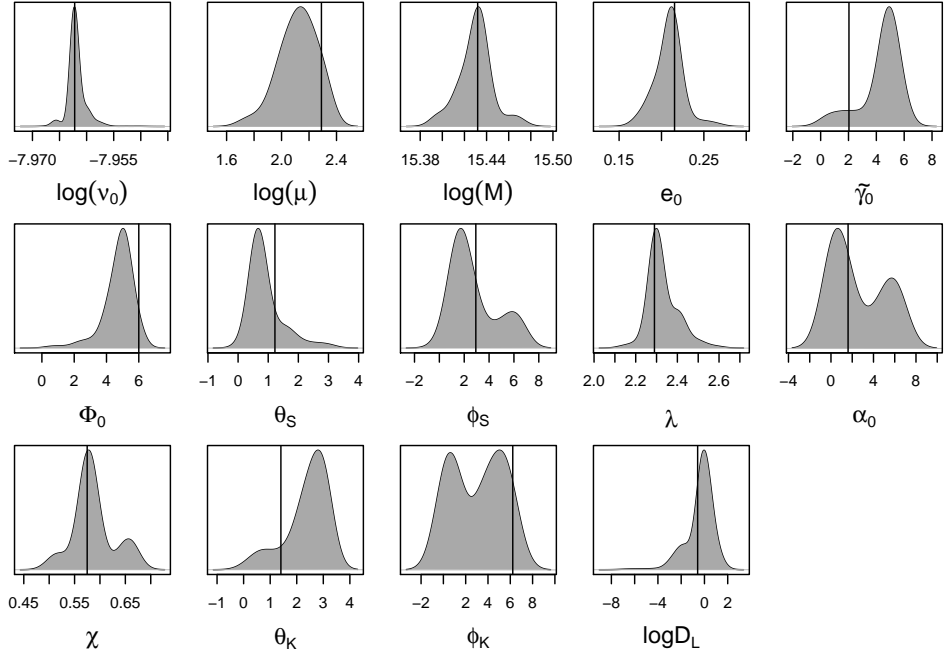} \centering
\caption{Kernel density estimates of the marginal posterior densities for all 14 parameters for the EMRI source MLDC 1B.3.2. The solid lines indicate the true values.
}
\label{Figure2}
\end{figure}

\begin{figure}[!h]
 \centering
 \includegraphics[width=1.00\textwidth]{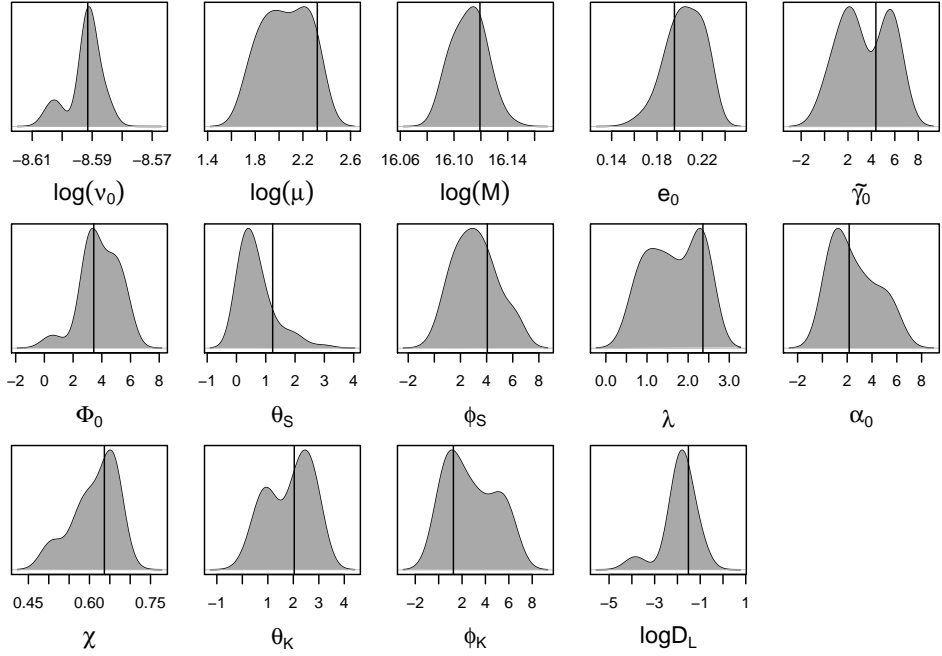} \centering
\caption{Kernel density estimates of the marginal posterior densities for all 14 parameters for the EMRI source MLDC 1C.3.1 (high mass source). The solid lines indicate the true values.}
\label{Figure3}
\end{figure}

In the simplest scenario the algorithm was applied to two EMRI data sets taken from earlier MLDC rounds; 1B.3.2 and 1C.3.1. 1B.3.2 contains a single medium mass EMRI source and 1C.3.1 contains a high mass EMRI source buried in the LISA instrument noise only. In these attempts the adjusted LFA was used and there was no problem in recovering the signal parameters using the true values as the starting points for the MCMC search, thus we used completely blind searches on two weeks long stretches of the data, to test the performance of our algorithm. In both cases the blind searches were conducted in multiple stages, i.e. first
several shorter MCMC chains were run in parallel without a swapping step from random starting points and those chains were chosen that showed stability and for which the SNR and likelihood values were larger than others. The modes of those chains were then set as the starting points in the next MCMC run. These steps were repeated for a few times to narrow down the search range. This approach is somewhat similar to that used in \cite{babak2009}. In the following, summaries of the posterior distribution of the parameters of the EMRI sources given in MLDC 1B.3.2 (medium mass) and MLDC 1C.3.1 (high mass) are given. Figure \ref{Figure1} display a typical shorter MCMC run demonstrating how the chains for different parameters find their true modes. Table \ref{Table1}  and figure \ref{Figure2} presents the posterior summary of the parameters for source 1B.3.2. Similarly, table \ref{Table2} and figure \ref{Figure3} show the results for source 1C.3.1. From these results it is clear that both signals were recovered with a great accuracy. The widths of different marginal posterior densities show that almost the whole prior range was searched before convergence.

\begin{table}[!h]
\caption{Posterior results and true parameter values for the recovered medium mass EMRI signal given in actual MLDC 1B.3.2 training data set.}
\scriptsize
\rm
\begin{tabular*}{\textwidth}{@{}l*{12}{@{\extracolsep{0pt plus12pt}}l}}
\br
Parameters & Mean & StdDev & Mode & 95\% BCI & True values & Prior range\\
\mr
$\log(\nu_{0})$      & -7.961883 & 0.001556 & -7.962237 & (-7.962645, -7.959406) & -7.962205 & $\log[5.0\times10^{-5},\; 0.01]$Hz\\
$\log(\mu)$          & 2.108143  & 0.137913 & 2.122627  & (1.853307, 2.323680)   & 2.290258  & $\log[9.5,10.5]M_{\odot}$\\
$\log(M)$            & 15.429324 & 0.014565 & 15.432732 & (15.401371, 15.460317) & 15.431952 & $\log[4.75,5.25]10^{6}M_{\odot}$\\
$e_{0}$              & 0.208397  & 0.015804 & 0.210424  & (0.183391, 0.229404)   & 0.215401  & $[0.10,0.40]$units\\
$\tilde{\gamma_{0}}$ & 4.384615  & 1.229447 & 4.845658  & (1.232442, 5.543105)   & 2.033297  & $[0,2\pi]$ rad\\
$\Phi_{0}$           & 4.682341  & 0.695824 & 5.011981  & (3.20136, 5.340058)    & 5.999900  & $[0,2\pi]$ rad\\
$\theta_{S}$         & 0.862168  & 0.531047 & 0.649613  & (0.492233, 2.114493)   & 1.222330  & $[0,\pi]$ rad\\
$\phi_{S}$           & 2.710713  & 1.905070 & 1.580461  & (1.092997, 6.237064)   & 2.934625  & $[0,2\pi]$ rad\\
$\lambda$            & 2.331862  & 0.068092 & 2.302875  & (2.251945, 2.463038)   & 2.289951  & $[0,\pi]$ rad\\
$\alpha_{0}$         & 2.884914  & 2.493591 & 0.649972  & (0.087803, 6.210357)   & 1.609215  & $[0,2\pi]$ rad\\
$\chi$               & 0.589667  & 0.039206 & 0.578192  & (0.515752, 0.65908)    & 0.574818  & $[0.5,0.7]$ $M^{2}$\\
$\theta_{K}$         & 2.478239  & 0.773029 & 2.870962  & (0.597917, 3.096335)   & 1.403416  & $[0,\pi]$ rad\\
$\phi_{K}$           & 3.373622  & 2.155152 & 5.050124  & (0.223237, 6.042101)   & 6.223129  & $[0,2\pi]$ rad\\
$D_{L}$              & -0.538350 & 1.128170 & -0.135892 & (-2.721751, 0.832533)  & -0.584778 & $\log[0,\infty]$ log(GPC)\\
\br
\end{tabular*}\label{Table1}
\end{table}     

\begin{table}
\caption{Posterior results and true parameter values for the recovered high mass EMRI signal given
in actual MLDC 1C.3.1 training data set.}
\scriptsize
\rm
\begin{tabular*}{\textwidth}{@{}l*{12}{@{\extracolsep{0pt plus12pt}}l}}
\br
Parameters & Mean & StdDev & Mode & 95\% BCI & True values & Prior range\\
\mr
$\log(\nu_{0})$ 	& -8.592 307 & 0.005 407 & -8.591 044 & (-8.604 332, -8.584 98)  & -8.591 472 & $\log[5.0\times10^{-5},0.01]$Hz\\
$\log(\mu)$ 		& 2.026 831  & 0.185 767 & 2.209 483  & (1.741 593, 2.322 630)   & 2.320 877  & $\log[9.5,10.5]M_{\odot}$\\
$\log(M)$ 		& 16.110 918 & 0.001 331 & 16.114 347 & (16.090 741, 16.132 173) & 16.119 304 & $\log[4.75,5.25]10^{6}M_{\odot}$\\
$e_{0}$ 		& 0.205 405  & 0.014 601 & 0.204 826  & (0.181 549, 0.226 693)   & 0.195 337  & $[0.10,0.40]$units\\
$\tilde{\gamma_{0}}$ 	& 3.425 854  & 2.011 777 & 2.116 807  & (0.189 037, 6.143 949)   & 4.381 526  & $[0,2\pi]$ rad\\
$\Phi_{0}$ 		& 3.948 827  & 1.273 087 & 3.366 273  & (0.966 292, 5.721 986)   & 3.441 184  & $[0,2\pi]$ rad\\
$\theta_{S}$ 		& 0.616 343  & 0.537 823 & 0.406 881  & (0.063 561, 1.823 876)   & 1.235 677  & $[0,\pi]$ rad\\
$\phi_{S}$ 		& 2.944 769  & 1.580 446 & 2.915 200  & (0.817 490, 6.224 024)   & 4.054 785  & $[0,2\pi]$ rad\\
$\lambda$ 		& 1.652 216  & 0.605 062 & 2.291 762  & (0.719 750, 2.352 924)   & 2.358 963  & $[0,\pi]$ rad\\
$\alpha_{0}$ 		& 2.610 697  & 1.849 872 & 1.261 552  & (0.361 536, 5.797 795)   & 2.158 356  & $[0,2\pi]$ rad\\
$\chi$ 			& 0.612 676  & 0.051 827 & 0.650 991  & (0.502 682, 0.662 555)   & 0.636 644  & $[0.5,0.7]$ $M^{2}$\\
$\theta_{K}$ 		& 1.887 538  & 0.871 457 & 2.464 749  & (0.472 588, 2.974 532)   & 2.036 360  & $[0,\pi]$ rad\\
$\phi_{K}$ 		& 2.827 086  & 2.093 720 & 1.124 890  & (0.152 617, 6.081 997)   & 1.260 128  & $[0,2\pi]$ rad\\
$D_{L}$ 		& -1.943 526 & 0.763 724 & -1.795 853 & (-3.843 257, -0.939 327) & -1.518 042 & $\log[0,\infty]$ log(GPC)\\
\br
\end{tabular*}\label{Table2}
\end{table}

\subsection{Multiple EMRIs}
4.2.1. MLDC 4 results. The approach was applied to detect signals generated by EMRI sources given in both training and blind data sets issued in the revised MLDC round 4. Looking at the amount of noise in these data, we attempted to recover signals from high mass EMRI systems only. Moreover, there were no medium mass sources in the training data set. Some preliminary results were presented at GWPAW (January 26\---29, 2011, Milwaukee, WI, USA). The MLDC notation for a high mass EMRI sources are EMRI-1-0, EMRI-1-1 and so on, and for a medium mass EMRI sources are EMRI-2-0, EMRI-2-1 and so on, whereas we denote the corresponding estimated sources (in the blind data) simply by High-0, High-1 and Med-0.

\subsubsection{Training data.}
\begin{figure}[!t]
 \centering
 \includegraphics[width=1.00\textwidth]{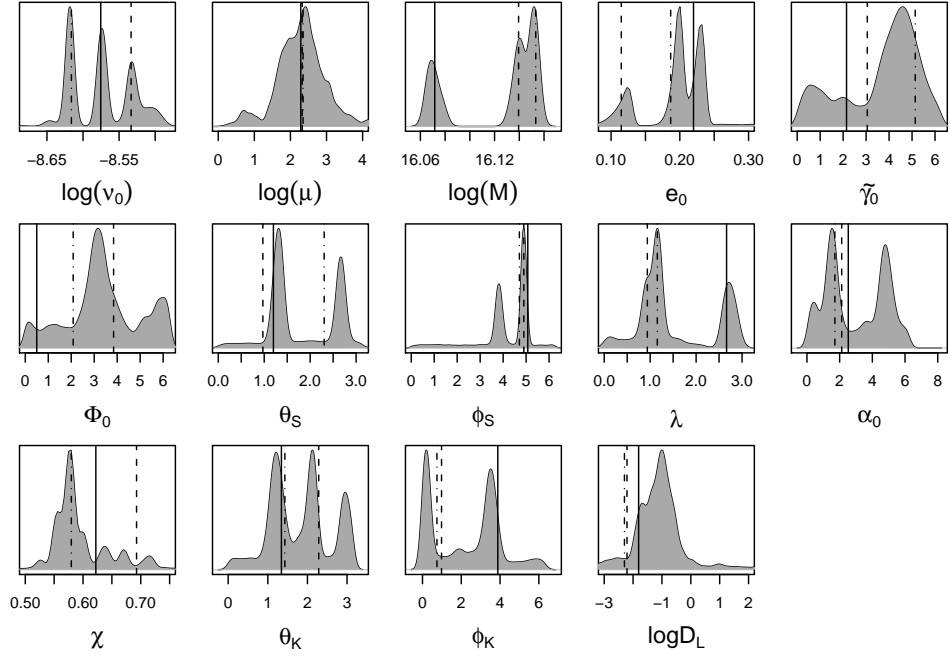} \centering
\caption{Kernel density estimates of the marginal posterior densities for all 14 parameters of
the first three EMRI sources in MLDC 4 training set. In these plots the solid lines indicate the
true parameter values of the source EMRI-1-0, whereas the dashed and dashed-dot-dashed lines
indicate the true parameter values for sources EMRI-1-1 and EMRI-1-2, respectively, in MLDC 4
training data.}
\label{Figure4}
\end{figure}

\begin{figure}[!h]
 \centering
 \includegraphics[width=1.00\textwidth]{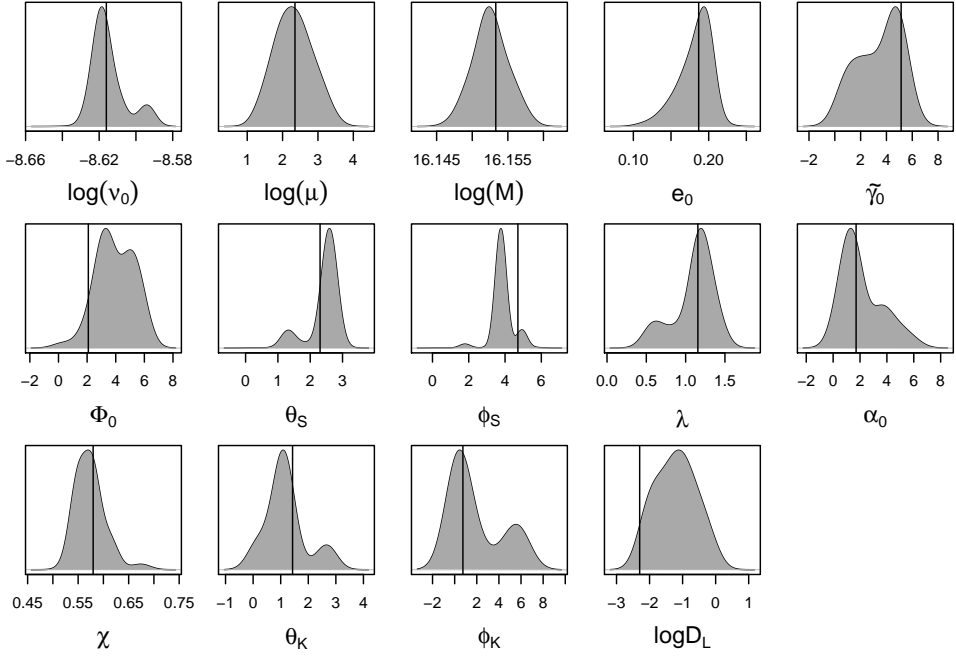} \centering
\caption{Kernel density estimates of the marginal posterior densities for all 14 parameters of the
third EMRI (EMRI-1-2) signal in MLDC 4 training data. The solid lines indicate the true parameter
values.}
\label{Figure5}

\end{figure}

The training data contains three high mass EMRI sources which are somewhat similar to each other, therefore a joint MCMC search was conducted to recover them. In an eight chain MCMC search on two weeks long data segments, three chains were started from the true parameter values of the three signals while the rest of the chains were started from the values in the vicinity of the true parameters. Figure \ref{Figure4} displays the results of this joint search. In the plots of kernel density estimates, different types of vertical lines denote the true values of the parameters of the three different high mass EMRI signals, named EMRI-1-0, EMRI-1-1 and EMRI-1-2 in MLDC 4 training keys, that can be found at the challenge webpage \cite{MLDC4}. After running for a sufficiently large number of iterations ($\sim4 \times 10^6$) it was observed that the third signal (dashed-dot-dashed vertical lines) was dominating the other two as can be seen in figure \ref{Figure4}, even though the overall mean swap acceptance rate between chains was $\sim35\%$. Thus the code was restarted with the starting values of different chains somewhat similar to the true parameter values of the third signal. These results are given in table \ref{Table3} and figure \ref{Figure5}. We can see that all the parameters, except the luminosity distance, $D_L$ , and some of the angles, are estimated with great accuracy and most of the parameters' chains show stability. From the MCMC trace plots the distance parameter seemed to be over estimated and the sky location ($\theta_S$ , $\phi_S$) seemed to have been locked at a different position. The wrong sky location was attributed to the fact that we were using LFA and are experiencing the same problem as was observed in \cite{gair2008} , which also used the LFA \cite{Barack}. At this stage, the problem of amplitude differences between the full LISA response and LFA was not known to us.

It was also evident that after some 800,000 iterations one of the neighbouring chains of the true ($T = 1$) chain found some other mode; however, overall the true chain was unaffected. This second mode could either correspond to, most probably, a low strength harmonic of this same EMRI signal as such a close harmonic usually corresponds to higher frequencies than the true one, or to a harmonic of another EMRI signal. This sort of overlapping and sharing of characteristics between different signals will be quite common in such complicated cases and will result in confusions among different EMRI signals.
\begin{table}
\caption{Posterior results for the third EMRI (EMRI-1-2) signal given in MLDC 4 training data
set.}
\scriptsize
\rm
\begin{tabular*}{\textwidth}{@{}l*{12}{@{\extracolsep{0pt plus12pt}}l}}
\br
Parameters 		& Mean 		& StdDev    & Mode 	 & 95\% BCI 			& True values& Prior range\\
\mr
$\log(\nu_{0})$ 	& -8.615 057 	& 0.008 435 & -8.618 036 & (-8.624 964, -8.593 739) 	& -8.616 096 & $\log[5.0\times10^{-5},0.01]$Hz\\
$\log(\mu)$ 		& 2.247 796 	& 0.519 295 & 2.185 567  & (1.433 541, 3.218 01)  	& 2.350 281  & $\log[9.5,10.5]M_{\odot}$\\
$\log(M)$ 		& 16.152 486 	& 0.002 285 & 16.152 319 & (16.148 756, 16.156 314) 	& 16.153 307 & $\log[4.75,5.25]10^{6}M_{\odot}$\\
$e_{0}$ 		& 0.184 634 	& 0.019 464 & 0.195 090  & (0.145 334, 0.203 263) 	& 0.186 712  & $[0.10,0.40]$units\\
$\tilde{\gamma_{0}}$ 	& 3.830 427 	& 1.407 267 & 4.734 815  & (1.089 336, 5.408 43) 	& 5.138 873  & $[0,2\pi]$ rad\\
$\Phi_{0}$ 		& 3.770 239 	& 1.079 385 & 3.244 351  & (2.128 266, 5.591 350) 	& 2.084 779  & $[0,2\pi]$ rad\\
$\theta_{S}$ 		& 2.455 871 	& 0.434 043 & 2.611 267  & (1.311 937, 2.755 807) 	& 2.305 033  & $[0,\pi]$ rad\\
$\phi_{S}$ 		& 3.901 775 	& 0.485 260 & 3.780 766  & (3.583 514, 4.948 959) 	& 4.707 928  & $[0,2\pi]$ rad\\
$\lambda$ 		& 1.115 781 	& 0.224 159 & 1.180 035  & (0.572 802, 1.395 177) 	& 1.155 677  & $[0,\pi]$ rad\\
$\alpha_{0}$ 		& 1.965 982 	& 1.351 007 & 1.372 157  & (0.521 520, 5.008 045) 	& 1.708 861  & $[0,2\pi]$ rad\\
$\chi$ 			& 0.573 573 	& 0.024 460 & 0.576 902  & (0.540 204, 0.616 836) 	& 0.579 723  & $[0.5,0.7]$ $M^{2}$\\
$\theta_{K}$ 		& 1.235 264 	& 0.685 402 & 1.127 951  & (0.096 748, 2.711 275) 	& 1.429 748  & $[0,\pi]$ rad\\
$\phi_{K}$ 		& 1.719 245 	& 2.280 232 & 0.380 733  & (0.088 012, 5.899 353) 	& 0.745 268  & $[0,2\pi]$ rad\\
$D_{L}$ 		& -1.262 741 	& 0.596 673 & -1.216 699 & (-2.187 479, -0.196 473) 	& -2.299 291 & $\log[0,\infty]$ log(GPC)\\
\br
\end{tabular*}\label{Table3}
\end{table}   

\subsubsection{Blind data.}
\begin{figure}[!t]
 \centering
 \includegraphics[width=1.00\textwidth]{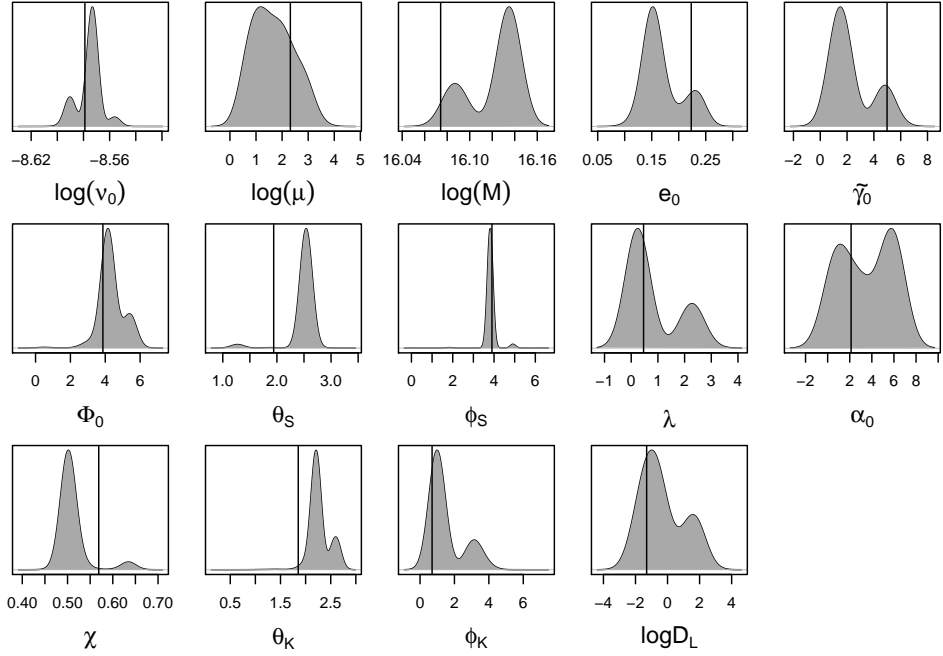} \centering
\caption{Kernel density estimates of the marginal posterior densities for all 14 parameters for the
detected EMRI source in MLDC 4 blind data. The solid lines indicate the true solutions of high
mass source EMRI-1-1 in MLDC 4 blind data.}
\label{Figure6}
\end{figure}

\begin{figure}[!t]
 \centering
 \includegraphics[width=1.00\textwidth]{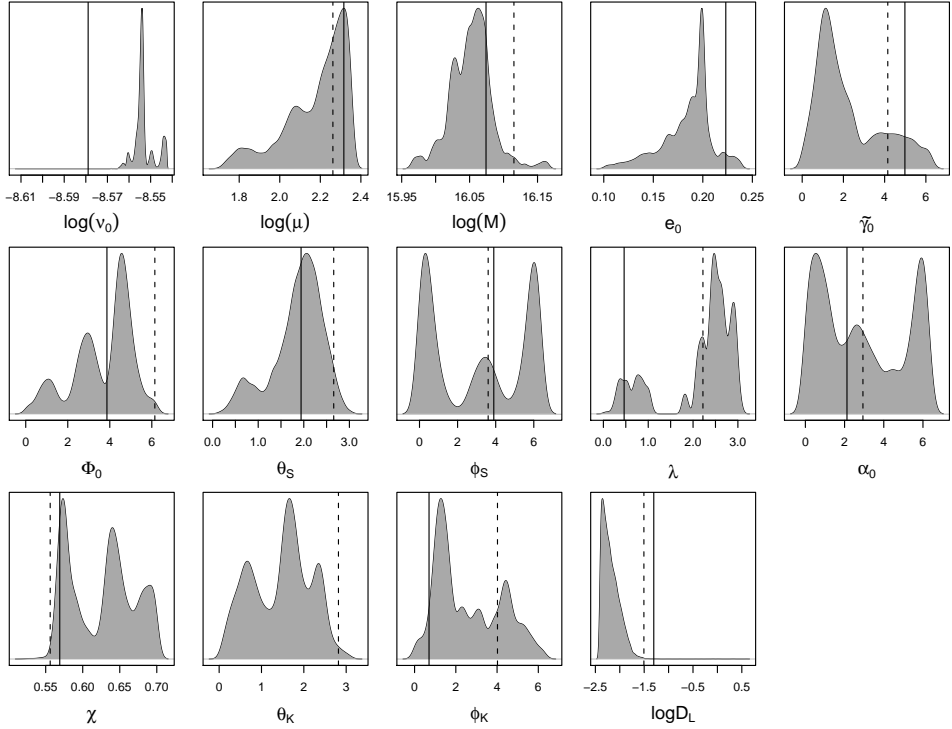} \centering
\caption{Kernel density estimates of the marginal posterior densities for all 14 parameters for the
high mass EMRI source `High-3' in MLDC 4 blind data. The dashed and solid lines corresponds
to the true solutions of two high mass sources EMRI-1-0 and EMRI-1-1, respectively, in MLDC 4
blind data.}
\label{Figure7}
\end{figure}

\begin{figure}[!t]
 \centering
 \includegraphics[width=0.80\textwidth]{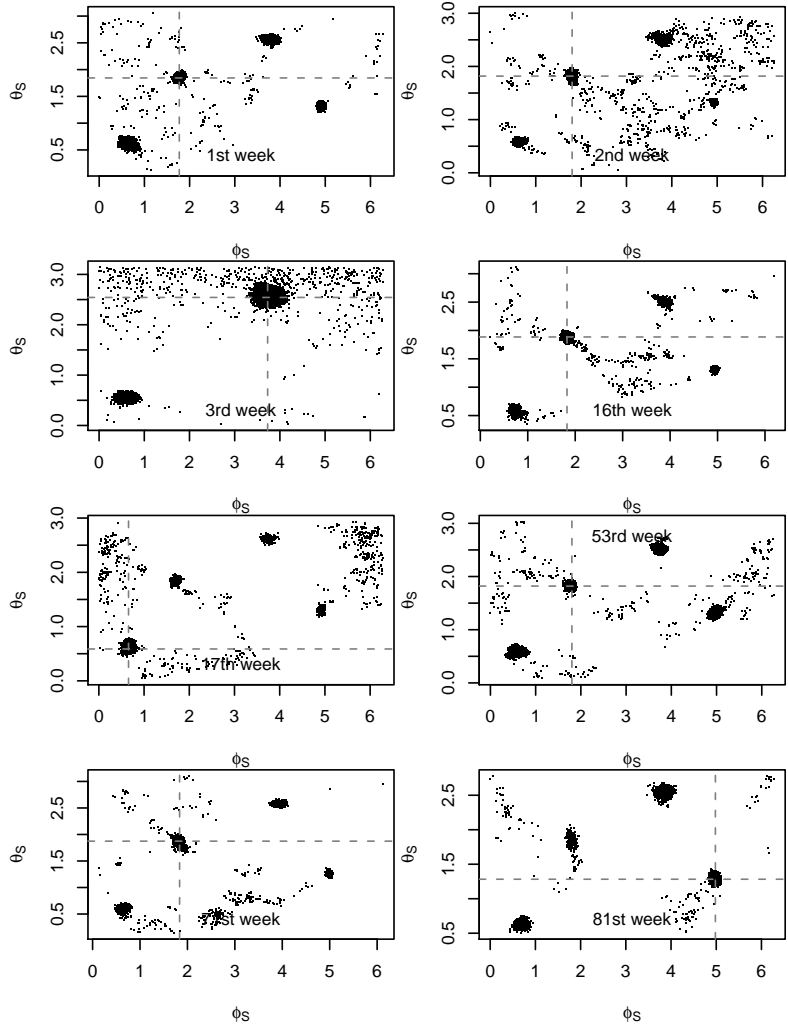} \centering
\caption{The joint plots of sky location angles for different time regions demonstrate that there are
four most probable sky positions either of the same source or there are two or more EMRI sources
located in different sky regions. The dashed lines indicate the \textit{maximum a posteriori} values.}
\label{Figure8}
\end{figure}

For the MLDC 4 blind data as a first step an eight chain MCMC search was conducted on the first two weeks in which all the eight chains were started from random values corresponding to a high mass EMRI source. Table \ref{Table4} shows a summary of the posterior estimates and figure \ref{Figure6} shows the kernel density estimates of the marginal posterior densities. The MCMC chains for all parameters showed great stability except for the CO's mass $\mu$ and distance $\log D_L$ the chains were showing somewhat oscillatory and correlated behaviour. 
The empirical correlation between these two parameters was $\sim0.92$, which is quite high. It could not be deciphered why this (high correlation) happened as such a phenomenon was found neither in earlier nor in subsequent searches. The chains for all the angles were also stable except for the parameter $\alpha_0$, which was vibrating between two different modes. An interesting result which was observed in our earlier searches, in which the LFA response was not adjusted (dividing by 3), is that for all time regions the joint plots of the sky location angles indicated a similar behaviour, though in the plots of the kernel density estimates of these two angles obtained for different time regions the (strongest) modes were different. For different time regions the joint plots of the two sky location angles are shown in figure \ref{Figure8}. These sky locations remained the same despite using different proposal distributions. At this stage we realized that amplitudes of LFA response are in general by a factor of $\sim3$ larger than full LISA response amplitudes. As an ad hoc solution to this problem, due to time limitation, the LFA amplitudes were divided by 3, which, in single source searches such as MLDC 1B (as shown in section 4.1), resulted not only in the correct estimation of distance and sky location parameters but also, in MLDC 4 searches, the sky location angles never got locked on wrong positions. 
\begin{table}[!t]
\caption{Posterior results for the detected high mass EMRI signal given in MLDC 4 blind data set
in the first step.}
\scriptsize
\rm
\begin{tabular*}{\textwidth}{@{}l*{12}{@{\extracolsep{0pt plus12pt}}l}}
\br
Parameters & Mean & StdDev & Mode & 95\% BCI & Prior range\\
\mr
$\log(\nu_{0})$      	& -8.573457 & 0.003015  & -8.573259 & (-8.574035, -8.572342)   & $\log[5.0\times10^{-5},0.01]$Hz\\
$\log(\mu)$          	& 1.409599  & 0.623259  & 1.179280  & (0.589129, 2.263096)     & $\log[9.5,10.5]M_{\odot}$\\
$\log(M)$            	& 16.133081 & 0.008141  & 16.134940 & (16.132126, 16.136896)   & $\log[4.75,5.25]10^{6}M_{\odot}$\\
$e_{0}$              	& 0.154880  & 0.014667  & 0.152115  & (0.139163, 0.166299)     & $[0.10,0.40]$units\\
$\tilde{\gamma_{0}}$ 	& 1.584359  & 0.708484  & 1.486492  & (0.859918, 2.269 012)    & $[0,2\pi]$ rad\\
$\Phi_{0}$           	& 4.197498  & 0.336251  & 4.147553  & (3.834165, 4.592633)     & $[0,2\pi]$ rad\\
$\theta_{S}$         	& 2.534345  & 0.098177  & 2.540198  & (2.507845, 2.579459)     & $[0,\pi]$ rad\\
$\phi_{S}$           	& 3.822423  & 0.097805  & 3.809877  & (3.744967, 3.886 976)    & $[0,2\pi]$ rad\\
$\lambda$            	& 0.298106  & 0.365414  & 0.249978  & (0.150734, 0.347221)     & $[0,\pi]$ rad\\
$\alpha_{0}$         	& 3.072884  & 2.491800  & 5.761360  & (0.150223, 6.189003)     & $[0,2\pi]$ rad\\
$\chi$               	& 0.499902  & 0.009253  & 0.501648  & (0.488 186, 0.508 558)   & $[0.5,0.7]$ $M^{2}$\\
$\theta_{K}$         	& 2.218992  & 0.075129  & 2.207951  & (2.161368, 2.274100)     & $[0,\pi]$ rad\\
$\phi_{K}$           	& 1.046394  & 0.370040  & 0.986633  & (0.870475, 1.121522)     & $[0,2\pi]$ rad\\
$D_{L}$              	& -0.928824 & 0.8035100 & -0.986087 & (-1.974076, -0.018514)   & $\log[0,\infty]$ log(GPC)\\
\br
\end{tabular*}\label{Table4}
\end{table}
In order to search the entire two years of the blind data several more MCMC searches were conducted on different time regions (data segments). In these searches we used a week long data segments chosen at regular gaps (generally six weeks) covering the entire two years duration of the data or till the anticipated plunge time of a typical EMRI signal (for most of the EMRI sources given in earlier rounds of MLDC the plunge generally occurs during $[1\frac{1}{2}\--1\frac{3}{4}]$ years, although in some cases the plunges are occurred in the last month of the two years duration). Each time region was searched individually by running eight or ten chain MCMCs. The search strategy was the same as explained in section 4.1. Attempts were made to recover both types of EMRI signals (high mass and medium mass) in these searches. In the end, for each type of EMRI source four possible best fits were chosen on the basis of high SNRs, to be submitted to MLDC 4. Among the submitted entries, the posterior results for a medium mass source `Med-2' and a high mass source `High-3' are presented in tables \ref{Table5} and \ref{Table6} and figure \ref{Figure7}. The rest of our entries and solutions can be found on MLDC 4 webpage \cite{MLDC4}. The kernel density estimates of the marginal posterior densities for most of the parameters show multi-modality. However, in most cases the strongest modes can be clearly recognized. In all these searches the average rates of the regular Metropolis acceptance were in 20\---40\% while the average swap acceptance rates were in 16\---30\%. The recovered medium EMRI source `Med-2', had a very high SNR but the estimated parameters are somewhat away from the true ones, demonstrating that these are probably the secondaries of the true densities. The `High-3'
has a lower SNR but the estimated parameters are quite close to the true ones.

\begin{table}
\caption{Posterior results for the detected EMRI signal `Med-2' given in MLDC 4 blind data set.}
\scriptsize
\rm
\begin{tabular*}{\textwidth}{@{}l*{12}{@{\extracolsep{0pt plus12pt}}l}}
\br
Source & Med-2 & SNR-A = 62.921 &  & SNR-E = 70.950 & SNR-AE = 103.708\\
\mr
Parameters & Mean & Mode & Std. dev. & Mean $\pm$ 2StdDev & Prior range\\
\mr
$\nu_{0}$ 		& 0.0003411537 	& 0.0003409980 	& 9.80e-7 & (0.0003390515, 0.0003429558) 	& $[5.0\times10^{-5},0.01]$Hz\\
$\mu$ 			& 13.03690058 	& 10.505164 49 	& 3.21e+0 & (6.603943913, 16.71099609) 		& $[9.5,10.5]M_{\odot}$\\
$M$ 			& 4840967.8525 	& 4838561.9866 	& 6.93e+4 & (4706125.2326, 4974725.6905) 	& $[4.75,5.25]10^{6}M_{\odot}$\\
$e_{0}$ 		& 0.218266 	& 0.217780 	& 2.23e-2 & (0.1731910, 0.2623679) 		& $[0.10,0.40]$units\\
$\tilde{\gamma_{0}}$ 	& 4.691678 	& 5.305615 	& 1.23e+0 & (2.8407054, 7.770524) 		& $[0,2\pi]$ rad\\
$\Phi_{0}$ 		& 5.176204 	& 4.961252 	& 8.86e-1 & (3.1897754, 6.732728) 		& $[0,2\pi]$ rad\\
$\theta_{S}$ 		& 1.382196 	& 1.408894 	& 2.88e-1 & (0.8333279, 1.984459) 		& $[0,\pi]$ rad\\
$\phi_{S}$ 		& 4.412142 	& 4.646086 	& 9.16e-1 & (2.8140819, 6.478091) 		& $[0,2\pi]$ rad\\
$\lambda$ 		& 0.694621 	& 0.681557 	& 2.78e-1 & (0.1265363, 1.236579) 		& $[0,\pi]$ rad\\
$\alpha_{0}$ 		& 2.624212 	& 0.261676 	& 2.68e+0 & (-5.101531, 5.624883) 		& $[0,2\pi]$ rad\\
$\chi$ 			& 0.581468 	& 0.578720 	& 2.12e-2 & (0.5362277, 0.621211) 		& $[0.5,0.7]$ $M^{2}$\\
$\theta_{K}$ 		& 1.588288 	& 1.562271 	& 3.12e-1 & (0.9389395, 2.185603) 		& $[0,\pi]$ rad\\
$\phi_{K}$ 		& 2.572490 	& 2.615519 	& 5.01e-1 & (1.6126596, 3.618377) 		& $[0,2\pi]$ rad\\
$D_{L}$ 		& 0.0792323 	& 0.0908401 	& 3.62e-2 & (0.01616134, 0.5105964) 		& $[0,\infty]$ (GPC)\\
\br
\end{tabular*}\label{Table5}
%
\caption{Posterior results for the detected EMRI signal `High-3' given in MLDC 4 blind data set.}
\scriptsize
\rm
\begin{tabular*}{\textwidth}{@{}l*{12}{@{\extracolsep{0pt plus12pt}}l}}
\br
Source & High-3 & SNR-A = 19.777 &  & SNR-E = 17.406 & SNR-AE = 26.346\\
\mr
Parameters & Mean & Mode & Std. dev. & Mean $\pm$ 2StdDev & Prior range\\
\mr
$\nu_{0}$ 		& 0.0001929449 	& 0.0001927831 	& 8.81e-7 & (0.0001910318, 0.0001945503) 	& $[5.0\times10^{-5},0.01]$Hz\\
$\mu$ 			& 8.81642159 	& 10.12691909 	& 1.24e+0 & (7.47961972, 13.71119041) 		& $[9.5,10.5]M_{\odot}$\\
$M$ 			& 9362002.4476 	& 9462096.5945 	& 2.86e+5 & (8901852.0130, 10057600.5800) 	& $[4.75,5.25]10^{6}M_{\odot}$\\
$e_{0}$ 		& 0.184721 	& 0.198859 	& 2.89e-2 & (0.148464, 0.249254) 		& $[0.10,0.40]$units\\
$\tilde{\gamma_{0}}$ 	& 2.247122 	& 1.133451 	& 1.59e+0 & (-2.049031, 4.315933) 		& $[0,2\pi]$ rad\\
$\Phi_{0}$ 		& 3.681986 	& 4.570893 	& 1.38e+0 & (1.791032, 7.350754) 		& $[0,2\pi]$ rad\\
$\theta_{S}$ 		& 1.864284 	& 2.0620089 	& 5.55e-1 & (0.952001, 3.172016) 		& $[0,\pi]$ rad\\
$\phi_{S}$ 		& 3.088018 	& 0.3096088 	& 2.44e+0 & (-4.581664, 5.20088177) 		& $[0,2\pi]$ rad\\
$\lambda$ 		& 2.147808 	& 2.4725720 	& 8.12e-1 & (0.848402, 4.09674241) 		& $[0,\pi]$ rad\\
$\alpha_{0}$ 		& 2.908103 	& 0.5411894 	& 2.13e+0 & (-3.719176, 4.80155457) 		& $[0,2\pi]$ rad\\
$\chi$ 			& 0.626321 	& 0.5732350 	& 4.31e-2 & (0.486933, 0.65953707) 		& $[0.5,0.7]$ $M^{2}$\\
$\theta_{K}$ 		& 1.466041 	& 1.6565712 	& 6.97e-1 & (0.262243, 3.05089978) 		& $[0,\pi]$ rad\\
$\phi_{K}$ 		& 2.621246 	& 1.2785729 	& 1.58e+0 & (-1.894009, 4.451155) 		& $[0,2\pi]$ rad\\
$D_{L}$ 		& 0.11290198 	& 0.09492333 	& 3.15e-2 & (0.0664814, 0.1355331) 		& $[0,\infty]$ (GPC)\\
\br
\end{tabular*}\label{Table6}
\end{table}
\subsection{Overlaps and SNRs}
\subsubsection{True signals: LFA versus full LISA response.}
When the true solutions for the MLDC 4 blind EMRIs were released, we tried to assess the performance of LFA and, in particular, the performance of the full AKW and TAKW in LFA regime by comparing the overlaps of their $A$ ($\equiv h_I$), $E$ ($\equiv h_{II}$) channels with those obtained with full LISA response. For this purpose all the true signals (noiseless) were generated using both lisatools package and our model codes. To calculate the overlap the following formula was used
\begin{equation}\label{equation11}
O = \dfrac{\langle a, b \rangle}{\sqrt{\langle a, a\rangle\;\langle b, b\rangle}},
\end{equation}
where $\langle a, b \rangle=\int_{-\infty}^{+\infty} \frac{a(f)b^*(f)}{S(f)}df$, the noise weighted inner product of two functions $a$, $b$. All
these quantities are in frequency domain. The lisatools package uses the full AKW model to generate the polarization signals ($h_+$, $h_\times$) of EMRIs and then employs the full LISA response function to generate TDI variables $X$, $Y$ and $Z$, that can then be transformed to $A$, $E$ channels. Similarly, our codes can generate $h_+$ and $h_\times$ polarizations using AKW or TAKW and then
compute $h_I$, $h_{II}$ using LFA. The overlap statistics are given in tables \ref{Table7} and \ref{Table8}. These results demonstrate that in the LFA regime the TAKW model gives better results than the full AKW. Therefore, with the LFA, it makes sense to use TAKW model for estimating the EMRI signals buried in LISA data.

\begin{table}
\caption{True signals: overlaps of the AKW and TAKW in LFA against AKW in full LISA response.}
\footnotesize
\rm
\begin{tabular*}{\textwidth}{@{}l*{12}{@{\extracolsep{0pt plus12pt}}r}}
\br
 & \multicolumn{3}{c}{AKW+LFA versus AKW+LISA} &  & \multicolumn{3}{c}{ TAKW+LFA versus AKW+LISA}\\
\cline{2-4} \cline{6-8} 
Signals          & EMRI-1-0 & \quad EMRI-1-1 & EMRI-2-0 &  & EMRI-1-0 & \quad EMRI-1-1 & EMRI-2-0\\
\mr
Overlap A        & 0.238    & 0.750    & 0.893    &  & 0.978    & 0.886    & 0.986\\
Overlap E        & 0.191    & 0.844    & 0.913    &  & 0.952    & 0.874    & 0.977\\
Combined overlap & 0.214    & 0.795    & 0.903    &  & 0.965    & 0.880    & 0.981\\
\br
\end{tabular*}\label{Table7}
%
\caption{Overlaps and SNRs of the detected signals for different sources.}
\footnotesize
\rm
{
\newcommand{\mc}[3]{\multicolumn{#1}{#2}{#3}}
\begin{tabular*}{\textwidth}{@{}l*{12}{@{\extracolsep{0pt plus12pt}}l}}
\br
 &  & Med-0$^\bigstar$ \qquad \quad& Med-1 \quad\qquad& Med-2$^\bigstar$ \quad\qquad & Med-3$^\bigstar$\quad\qquad\\
\mr
EMRI-2-0$^\bigstar$ \qquad& Overlap A & \mc{1}{c}{0.501}  & \mc{1}{c}{0.102} & \mc{1}{c}{0.219} & \mc{1}{c}{0.185}\\
 & Overlap E         & \mc{1}{c}{0.664}  & \mc{1}{c}{0.161} & \mc{1}{c}{0.134} & \mc{1}{c}{0.115}\\
 & Combined overlap  & \mc{1}{c}{0.577}  & \mc{1}{c}{0.128} & \mc{1}{c}{0.171} & \mc{1}{c}{0.146}\\
 & SNR A             & \mc{1}{c}{38.133} & \mc{1}{c}{1.647} & \mc{1}{c}{65.307} & \mc{1}{c}{62.921}\\
 & SNR E             & \mc{1}{c}{42.003} & \mc{1}{c}{1.190} & \mc{1}{c}{80.562} & \mc{1}{c}{70.950}\\
 & Combined SNR      & \mc{1}{c}{56.731} & \mc{1}{c}{2.032} & \mc{1}{c}{103.708} & \mc{1}{c}{94.831}\\
 &                   & High-0 & High-1$^\blacklozenge$ & High-2 & High-3$^\clubsuit$\\
EMRI-1-0$^\clubsuit$ & Overlap A & \mc{1}{c}{0.021}  & \mc{1}{c}{0.114}  & \mc{1}{c}{0.017} & \mc{1}{c}{0.055}\\
 & Overlap E         & \mc{1}{c}{0.018}  & \mc{1}{c}{0.084}  & \mc{1}{c}{0.041} & \mc{1}{c}{0.125}\\
 & Combined overlap  & \mc{1}{c}{0.019}  & \mc{1}{c}{0.098}  & \mc{1}{c}{0.027} & \mc{1}{c}{0.083}\\
EMRI-1-1$^\blacklozenge$ & Overlap A & \mc{1}{c}{0.018}  & \mc{1}{c}{0.199}  & \mc{1}{c}{0.018} & \mc{1}{c}{0.031}\\
 & Overlap E         & \mc{1}{c}{0.007}  & \mc{1}{c}{0.286}  & \mc{1}{c}{0.024} & \mc{1}{c}{0.057}\\
 & Combined overlap  & \mc{1}{c}{0.012}  & \mc{1}{c}{0.238}  & \mc{1}{c}{0.021} & \mc{1}{c}{0.042}\\
 & SNR A             & \mc{1}{c}{11.173} & \mc{1}{c}{11.262} & \mc{1}{c}{2.688} & \mc{1}{c}{19.777}\\
 & SNR E             & \mc{1}{c}{15.589} & \mc{1}{c}{10.227} & \mc{1}{c}{3.303} & \mc{1}{c}{17.406}\\
 & Combined SNR      & \mc{1}{c}{19.180} & \mc{1}{c}{15.213} & \mc{1}{c}{4.258} & \mc{1}{c}{26.346}\\
\br
\end{tabular*}
}
\label{Table8}
\end{table}
\subsubsection{True versus detected signals.}
Overlaps of the estimated signals (TAKW + LFA) with true signals (AKW + full LISA response) and their SNRs are shown in table \ref{Table8}. Different symbols indicate the correspondence between the best matches of estimated signals with true signals. The estimated medium mass signal `Med-2' is believed to have a best match with the only medium mass source in MLDC 4 blind data. The low degree of overlaps and SNRs of the estimated high mass signals is due to the fact that for these sources the difference factor between the LFA and full LISA response is $\sim6$ rather than $\sim3$.

\section{Conclusion}
In this paper we report on the application of a Bayesian approach to statistical inference on the highly challenging problem of detecting and estimating parameters of GW signals produced by EMRI's in the data from a future space-based interferometric mission such as LISA. The LFA is used to model the LISA response and the Whittle likelihood as a realistic approximation to the unknown error distribution. Differences in the amplitudes of LFA and the LISA response function were found and temporarily fixed. Further investigations are required to overcome these discrepancies. We assume the noise power spectrum to be unknown and formulate an Inv$-\chi^2$ posterior with a conjugate prior distribution for the unknown noise parameters. We have implemented a parallel tempering MCMC algorithm to sample from the posterior distribution of all 14 unknown parameters, the first instance of a fully Markovian algorithm for EMRI detection and characterization in the context of MLDC LISA data analysis challenges. This algorithm has the potential to sample from multi-modal distributions and thus to avoid getting trapped in local modes. Our simulation results using single as well as multiple EMRI sources in MLDC training and blind data should be seen as a snapshot of an investigation still in progress. However, they show that this strategy holds great promise for posterior computation under more realistic conditions and hints at areas where the algorithm could be refined and tuned.
\ack
AA acknowledges support from the Higher Education Commission of Pakistan. The work of NC was partially supported by NSF grant PHY-0854790. We also thank the Max-Planck-Institut f\"{u}r Gravitationsphysik (Albert-Einstein-Institut), Hannover, Germany, for access to their computing facilities.

\section*{References}
  \bibliographystyle{unsrt}
  \bibliography{bibliography}

\end{document}